\documentclass[final,english]{bullsrsl}[2022/06/15]

\usepackage[latin1,utf8]{inputenc}
\usepackage[T1]{fontenc}

\usepackage{natbib} 
\usepackage{graphicx}

\def\lya{Ly$\alpha$ }

\begin{document}
\title{GraL spectroscopic identification of multiply imaged quasars}
\author[affil={1}, corresponding]{Priyanka}{Jalan}
\author[affil={2}]{Vibhore}{Negi}
\author[affil={3}]{Jean}{Surdej}
\author[affil={4}]{C\'eline}{Boehm}
\author[affil={3}]{Ludovic}{Delchambre}
\author[affil={5}]{Jakob Sebastian}{den Brok}
\author[affil={6}]{Dougal}{Dobie}
\author[affil={7}]{Andrew}{Drake}
\author[affil={8}]{Christine}{Ducourant}
\author[affil={7}]{S. George}{Djorgovski}
\author[affil={9}]{Laurent}{Galluccio}
\author[affil={7}]{Matthew J.}{Graham}
\author[affil={10,11}]{Jonas}{Kl\"uter}
\author[affil={12, 13}]{Alberto}{Krone-Martins}
\author[affil={8}]{Jean-Fran{\c c}ois}{LeCampion}
\author[affil={7}]{Ashish A.}{Mahabal}
\author[affil={9}]{Fran{\c c}ois}{Mignard}
\author[affil={6}]{Tara}{Murphy}
\author[affil={14}]{Anna}{Nierenberg}
\author[affil={15}]{Sergio}{Scarano}
\author[affil={14}]{Joseph}{Simon}
\author[affil={9}]{Eric}{Slezak}
\author[affil={3}]{Dominique}{Sluse}
\author[affil={16}]{Carolina}{Sp\'indola-Duarte}
\author[affil={14}]{Daniel}{Stern}
\author[affil={16}]{Ramachrisna}{Teixera}
\author[affil={10,17}]{Joachim}{Wambsganss}


\affiliation[1]{Center for Theoretical Physics, Polish Academy of Sciences, al. Lotnik\'{o}w 32/46, 02-668 Warsaw, Poland}
\affiliation[2]{Aryabhatta Research Institute of Observational Sciences, Nainital, Uttarakhand, India}
\affiliation[3]{Institut d'Astrophysique et de G\'{e}ophysique, Universit\'{e} de Li\`{e}ge, All\'{e}e du 6 Ao\^{u}t 19c, B-4000 Li\`{e}ge, Belgium}
\affiliation[4]{School of Physics, The University of Sydney, NSW 2006, Australia}
\affiliation[5]{Sydney Institute for Astronomy, School of Physics, University of Sydney, NSW 2006, Australia}
\affiliation[6]{Argelander-Institut f\"{u}r Astronomie, Universit\"{a}t Bonn, Auf dem H\"{u}gel 71, D-53121 Bonn, Germany}
\affiliation[7]{Cahill Center for Astronomy and Astrophysics, California Institute of Technology, 1216 E.  California Blvd., Pasadena, CA 91125, US}
\affiliation[8]{Laboratoire d'Astrophysique de Bordeaux, Univ. Bordeaux, CNRS, B18N, all\'{e}e Geoffroy Saint-Hilaire, 33615 Pessac, France}
\affiliation[9]{Universit\'{e} C\^{o}te d'Azur, Observatoire de la C\^{o}te d'Azur, CNRS, Laboratoire Lagrange, Boulevard de l'Observatoire, CS 34229, 06304 Nice, France}
\affiliation[10]{Zentrum f\"{u}r Astronomie der Universit\"{t} Heidelberg, Astronomisches Rechen-Institut, M\"{o}nchhofstr. 12-14, 69120 Heidelberg, Germany}
\affiliation[11]{Department of Physics \& Astronomy, Louisiana State University, 261 Nicholson Hall, Tower Dr., Baton Rouge, LA 70803-4001, US}
\affiliation[12]{Donald Bren School of Information and Computer Sciences, University of California, Irvine, Irvine CA 92697, US}
\affiliation[13]{CENTRA, Faculdade de Cincias, Universidade de Lisboa, 1749-016, Lisbon, Portugal}
\affiliation[14]{Jet Propulsion Laboratory, California Institute of Technology, 4800 Oak Grove Drive, Mail Stop 264-789, Pasadena, CA 91109, USA}
\affiliation[15]{Departamento de F\'{i}sica - CCET, Universidade Federal de Sergipe, Rod. Marechal Rondon s/n, 49.100-000, Jardim Rosa Elze, S\~{a}o Crist\'{o}v\~{a}o, SE, Brazil}
\affiliation[16]{Instituto de Astronomia, Geof\'isica e Ci\^encias Atmosf\'ericas, Universidade de S\~{a}o Paulo, Rua do Mat\~{a}o, 1226, Cidade Universit\'aria, 05508-900 S\~{a}o Paulo, SP, Brazil}
\affiliation[17]{International Space Science Institute (ISSI), Hallerstra\ss e 6, 3012 Bern, Switzerland}

\correspondance{priyajalan14@gmail.com}
\date{17 May 2023}

\maketitle

\begin{abstract}
Gravitational lensing is proven to be one of the most efficient tools for studying the Universe. The spectral confirmation of such sources requires extensive calibration. This paper discusses the spectral extraction technique for the case of multiple source spectra being very near each other. Using the \emph{masking technique}, we first detect high Signal-to-Noise (S/N) peaks in the CCD spectral image corresponding to the location of the source spectra. This technique computes the cumulative signal using a weighted sum, yielding a reliable approximation for the total counts contributed by each source spectrum. We then proceed with the subtraction of the contaminating spectra. Applying this method, we confirm the nature of 11 lensed quasar candidates. 
\end{abstract}


\section{Introduction}
\label{s:ch6intro}

The two chief methods for estimating the Hubble-Lema{\^i}tre constant ($H_o$) are (i) studying the relation between the distances and redshifts of the objects in the  Universe \citep[see, e.g.,][]{Lemaitre1927ASSB...47...49L, Hubble1929PNAS...15..168H}, (ii) interpolating the expansion rate from models based on the Cosmic Microwave Background Radiation \citep[see, e.g.,][]{Efstathiou1990Natur.348..705E, Raul2003ApJ...593..622J} and Type-1a supernovae \citep[see, e.g.,][]{Riess1998AJ....116.1009R, Perlmutter1999ApJ...517..565P}. However, studying gravitational lens systems has proven to be a complementary tool to determine $H_o$ in a model-independent manner \citep[see, e.g.,][]{Refsdal1964MNRAS.128..307R, Blandford1992ARA&A..30..311B, Surdej1994ASSL..187..409S, Patrick2023}. 

\begin{figure}
\centering
\includegraphics[width=0.75\textwidth]{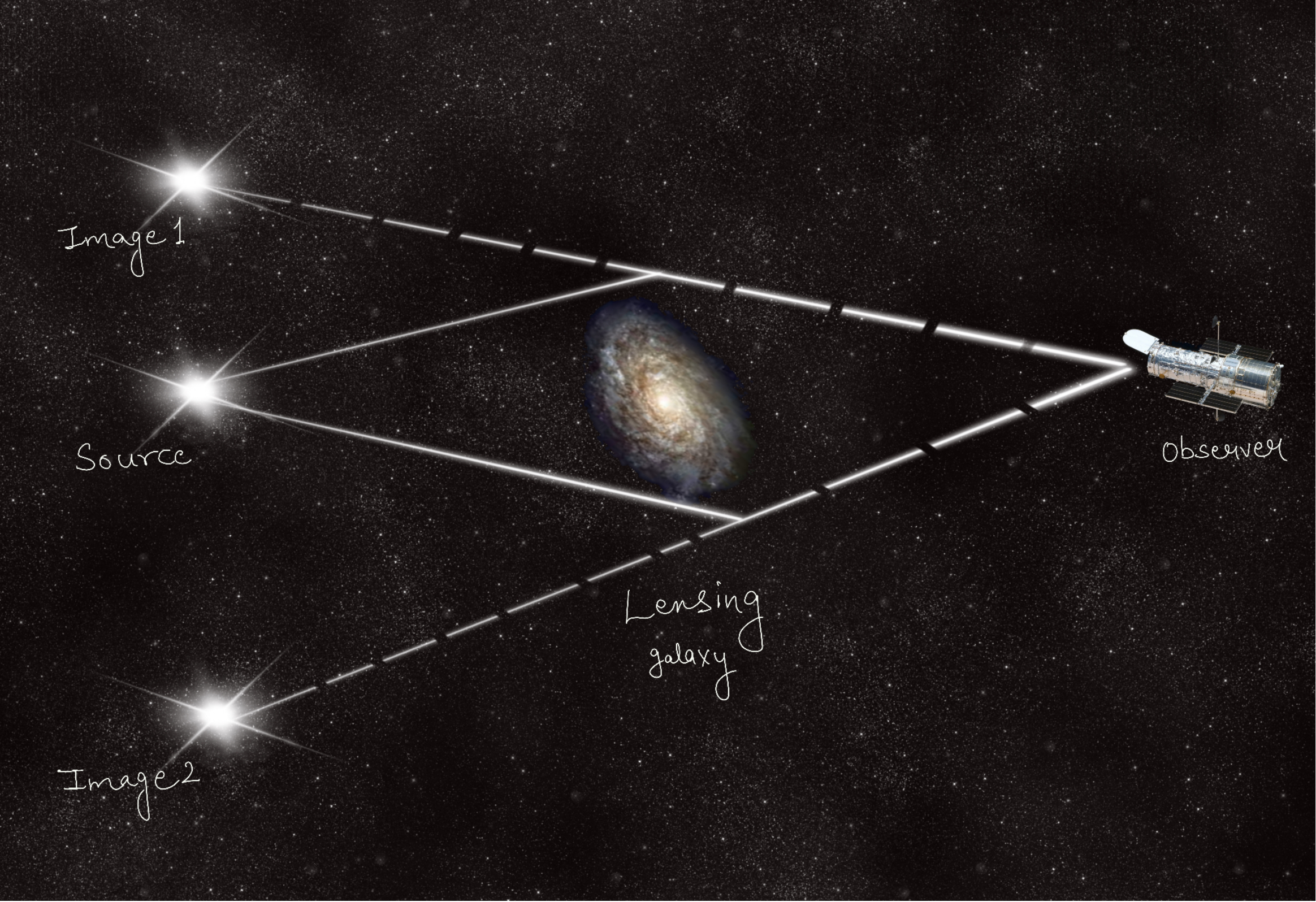}
\bigskip

\begin{minipage}{12cm}
\caption{This artistic image shows light deflection from a background quasar due to a foreground galaxy producing two lensed images of the quasar.} 
\label{fig:lens}
\end{minipage}
\end{figure}

Gravitational lensing is due to the deflection of light from a background source caused by a massive foreground structure (see, e.g., Fig.\,\ref{fig:lens}). The mass of the foreground source leads to a curved space-time that bends the light travelling from the background source to the observer. The deflection of light depends on the distribution of mass within the foreground structure and the distances between the observer, the lens and the source. Such lensing can produce multiple images and/or amplification of the background source. The delay between the light travel times corresponding to the various lensed images contains information about $H_o$. In addition to the estimation of $H_o$, gravitational lens systems can also be used to evaluate the mass of the foreground structure as well as properties of dark energy and matter \citep[see, e.g.,][and references therein]{Cao2015ApJ...806..185C}. Furthermore, the transverse correlation of the \lya forest clouds in the IGM can also be studied using gravitationally lensed quasars \citep[][]{Bechtold1995AJ....110.1984B, Dolan2000ApJ...539..111D, Rauch2001ApJ...562...76R, 2003MNRAS.340..937T}.
\par The concept of gravitational lensing is traced back to 1704 when Newton, in his work {\it Opticks}, questions whether objects may interact with light at a distance and bend the light beams. It was in 1784 that Henry Cavendish, using Newton's gravitational theory and Newtonian optics, computed the deflection angle of light due to a point mass, where light is supposed to be composed of corpuscles being affected by a gravitational field in the same way as material particles \citep[detailed in][]{Lotze2022AnP...53400102L}. Later, Johann Georg von Soldner calculated the angle of deflection of light passing at a distance `b' from a mass M as,
\begin{equation}
\alpha_N = \frac{2GM}{c^2b}.
\end{equation}

here, G is the gravitational constant, and c is the speed of light in a vacuum.
\par In 1915, Albert Einstein calculated the deflection angle of light by the Sun using his general theory of relativity and found that this angle was twice that previously predicted. In 1937, Fritz Zwicky \citep[][]{Zwicky1937PhRv...51..290Z} argued that this phenomenon might let galaxy clusters serve as gravitational lenses.
It was not until 1979 that the detection of the twin quasars SBS 0957+561 by \citet[] []{Walsh1979Natur.279..381W} was shown to be the first example of a doubly imaged quasar.
\par The discovery of such exotic sources requires overcoming various observational challenges. Not only are these objects rare, but they are also challenging to be identified in large catalogues. Therefore, only two hundreds of spectroscopically confirmed gravitational systems (\url{https://research.ast.cam.ac.uk/lensedquasars/index.html}) are currently known. Recent studies have tried discovering these systems using machine learning algorithms and large astronomical surveys. One such effort is made by the Gaia-GraL (Gaia Gravitational Lens systems) group using the Gaia data, an all-sky survey space-based mission designed to catalogue the stars in the Milky Way. However, while doing so, various extragalactic sources are also unveiled in this process \citep[][]{Tsalmantza2012A&A...537A..42T, Krone2013A&A...556A.102K,  Souza2014A&A...568A.124D, Delchambre2018MNRAS.473.1785D,  Bailer2019MNRAS.490.5615B, Creevey2023A&A...674A..26C}. The best angular resolution of 0.18$^{\prime \prime}$ achievable with Gaia does provide an excellent opportunity to search for gravitational lens candidates \citep[][]{Agnello10.1093/mnras/sty1419, Lemon10.1093/mnras/sty911}. \citet[][]{Finet2016A&A...590A..42F} have predicted to find $\sim$ 2900 such lensed quasar candidates in the Gaia survey. Since then the Gaia-GraL team has spectroscopically confirmed 15 quadruply imaged systems and seven doubly ones \citep[see, e.g.,][]{GraL1_2018A&A...616L..11K, GraL2_2018A&A...618A..56D, GraL3_2019A&A...622A.165D, GraL4_2019A&A...628A..17W, GraL5_2019arXiv191208977K, GraL6_2020arXiv201210051S}. 
\par In this paper, we discuss one of the data reduction techniques developed to spectroscopically confirm these sources to be gravitationally lensed quasars.

\section{Observations}
\label{s:obs_ch6}
The spectroscopic observations of the gravitationally lensed candidates were performed using EFOSC2 installed at the New Technology Telescope (\url{https://www.eso.org/sci/facilities/lasilla/instruments/efosc.html}) and ADFOSC equipping the 3.6 m Devasthal Optical Telescope (\url{https://www.aries.res.in/facilities/astronomical-telescopes/360cm-telescope/Instruments}). 

\section{Spectroscopy}
\label{s:spectro}
\begin{figure} 
\centering
\includegraphics[width=0.75\textwidth]{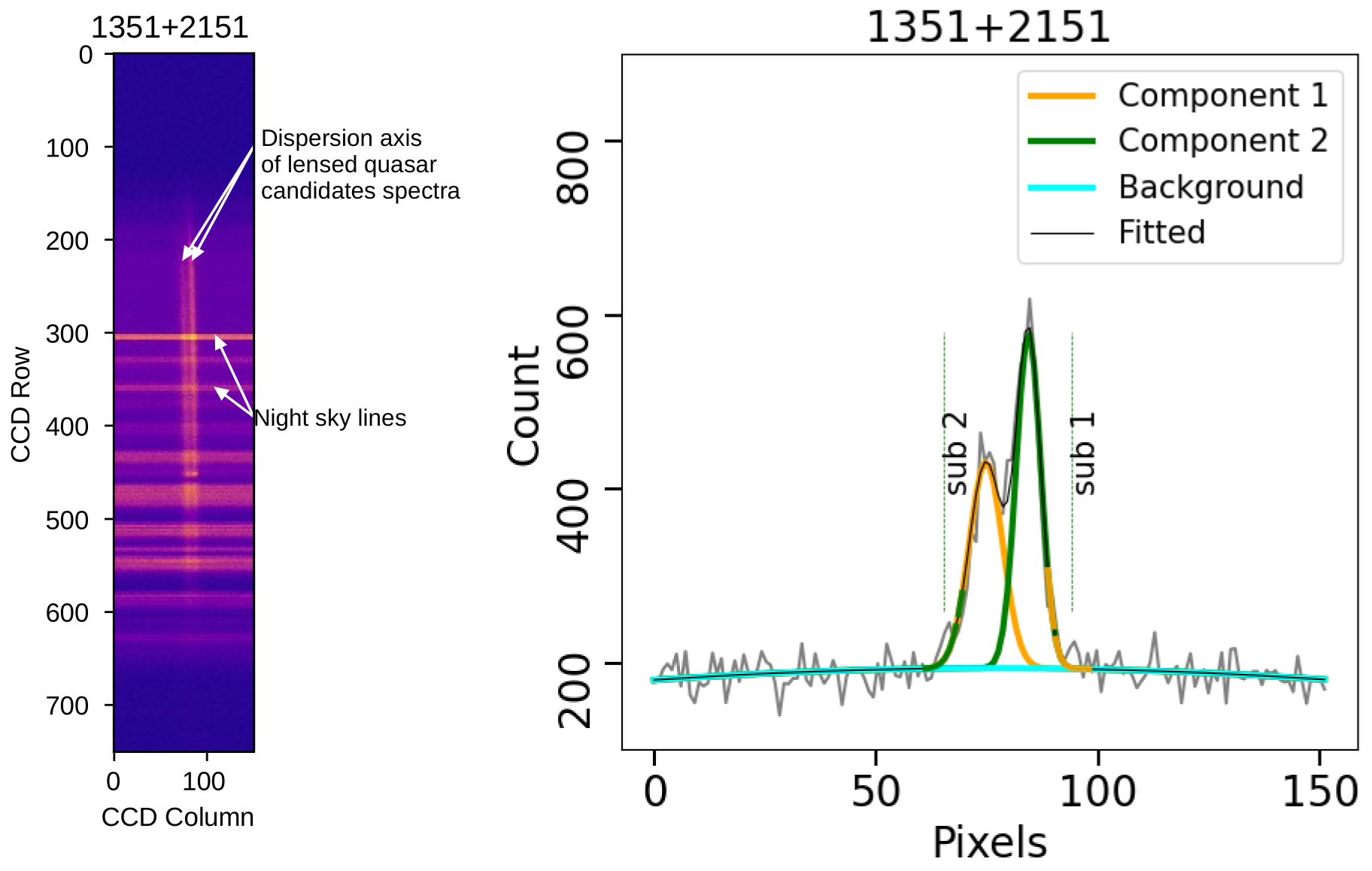}
\bigskip

\begin{minipage}{12cm}
\caption{{\it Left:} CCD frame highlighting the spectra of two nearby lensed quasar images and superimposed night skylines. {\it Right:} Slice of the CCD along row number 270 shows the superposition of two almost Gaussian profiles. In Section\,\ref{s:decont}, the technique to remove the contamination of neighbouring sources is detailed.}
\label{Fig:decontam}
\end{minipage}
\end{figure}

The extraction process that uses the {\tt apall} task in IRAF is quite simple if the two nearby source spectra are well separated. However, for most of the gravitational lens systems, the multiple lensed quasar images are very near each other, complicating the extraction process of individual non-contaminated spectra (e.g., left panel of Fig.\,\ref{Fig:decontam}). When extracting the spectrum of a designated source component, it is important to subtract the contamination due to all the other ones correctly. The right panel of Fig.\,\ref{Fig:decontam} shows counts versus pixel numbers along a fixed row of the CCD spectral image illustrated in the left panel. Firstly, we assume a {\it Gaussian distribution} of the photon counts from a source for a particular CCD row, i.e., the point spread function (PSF). Therefore, we have over-plotted two Gaussian profiles associated with the two nearby spectra (say due to component  C$_1$ and component C$_2$) along with the background in cyan colour. In the right panel of Fig.\,\ref{Fig:decontam}, the red colour represents the contaminated area by the neighbouring source. The region of C$_2$ contaminating the flux of C$_1$ is similar to sub$_1$ (shown in orange dashed lines). Similarly, the region of C$_1$ contaminating the flux of C$_2$ is similar to sub$_2$ (shown in green dashed lines). So in order to retain uncontaminated spectra,
\begin{equation}
  \begin{array}
    cC_1^{decont} = C_1^{cont} -sub_1 ~~;~~
    C_2^{decont} = C_2^{cont} -sub_2.
  \end{array}
\end{equation}

The extraction using IRAF gives good results for bright and well-separated sources; however, faint and very nearby sources require a more intricate extraction technique. 

 \subsection{Decontamination of nearby spectra}
 \label{s:decont}
{ \bf Masking technique:} The optimal spectral extraction that we propose makes use of a \emph{masking technique}, as briefly discussed below. This technique is advantageous in the case of doubly imaged quasars for which both spectra lie very close to each other and significantly contaminate each other. Also, this algorithm is highly useful in reducing the statistical noise of the extracted spectra by assigning non-uniform pixel weights during the extraction.
  
\begin{figure}
\centering
\includegraphics[width=0.5\textwidth]{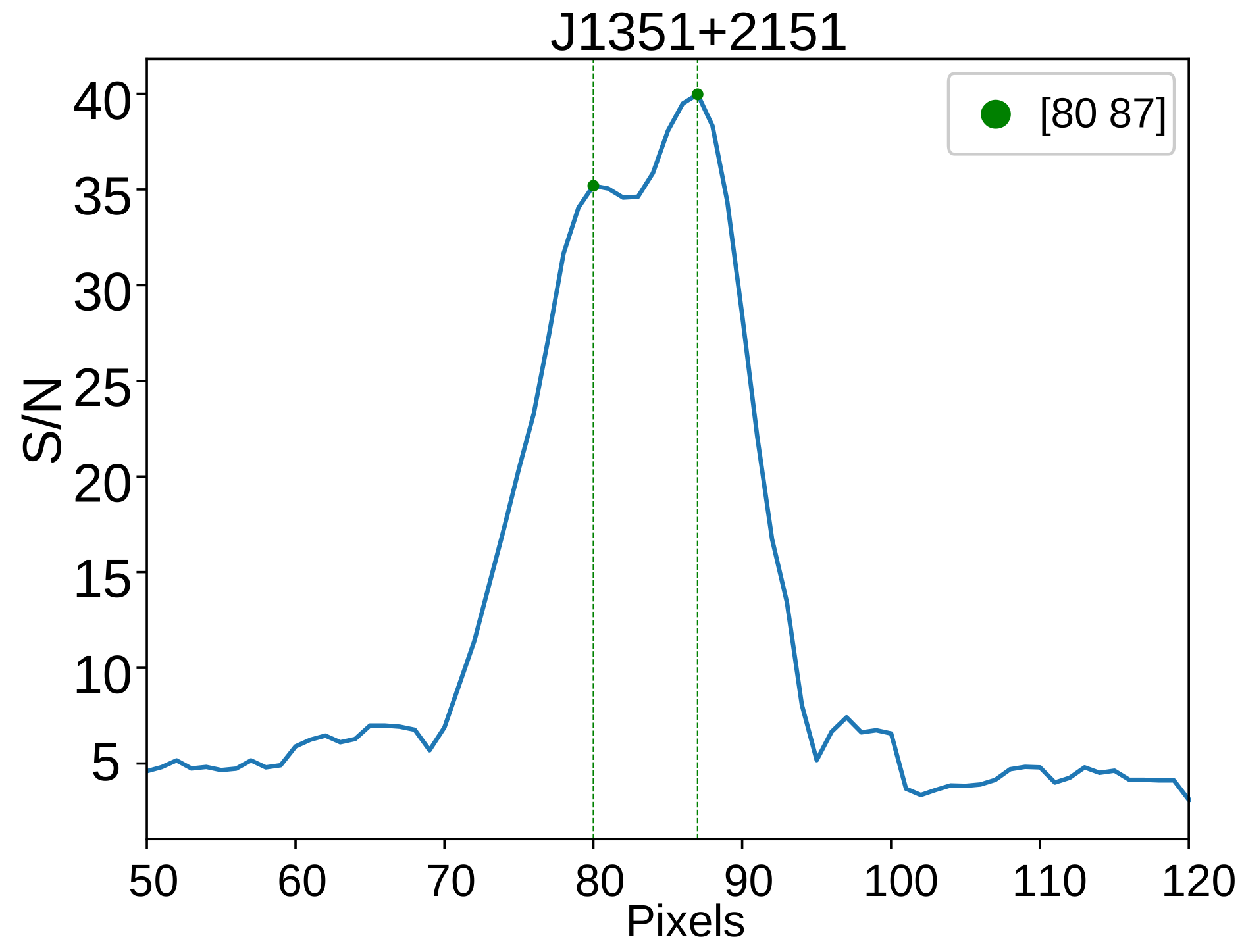}
\bigskip

\begin{minipage}{12cm}
\caption{S/N versus pixel number for the CCD image shown in the left panel of Fig.\,\ref{Fig:decontam}.}
\label{Fig:mask}
\end{minipage}
\end{figure}

\begin{figure}
\centering
\includegraphics[width=\textwidth]{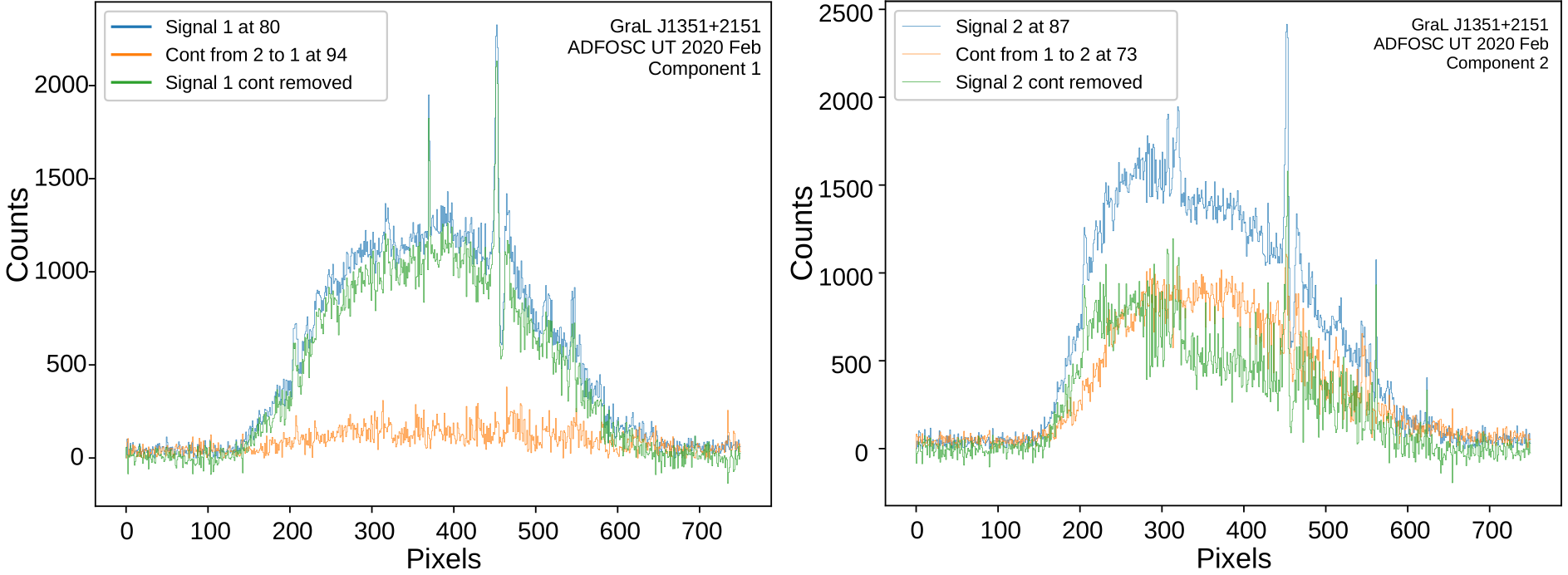}
\smallskip

\begin{minipage}{12cm}
    \caption{The blue spectra are those  extracted at positions $X_1=80$ and $X_2=87$  from the spectral CCD frame shown in
      Fig.\,\ref{Fig:decontam}. The flux from the contamination at  positions $X_1-(X_2-X_1)$ and at positions $X_2 + (X_2-X_1)$ are orange. The decontaminated spectra $I_D[X_1] = I[X_1=80] - I[X_2+(X_2 - X_1)=94]$ and $I_D[X_2] = I[X_2=87] - I[X_1-(X_2 - X_1)=73]$ are shown in green colour.}
      \label{Fig:mask_res}
\end{minipage}
\end{figure}

\begin{enumerate}
  \renewcommand*\labelenumi{[\theenumi]}
\item After pre-processing the science image, the CCD image $(I_*+Sky)$, as shown in the left panel of Fig.\,\ref{Fig:decontam}, is used to create several other images (see below).
\item The background image ($Sky$) is created after applying a sigma clipping algorithm to the spectra along each row. The sky array is the median of the remaining pixels along each row. We then subtract from each column this median sky background.
\item We then generate a binary mask, which in the case of a single source, would mimic the shape of the displayed source spectrum. We assign the value of 1 to each pixel inside the mask and 0 everywhere else. In the case of a double source spectrum, we construct the mask by fitting, at best, its shape to that of the brightest source spectrum.  
\item The mask is then multiplied with the science image step-[1] generating the $(I_*+Sky)_M$ masked image (the subscript $M$ indicates that the frame has been masked).
\item We also multiply the mask with the background subtracted image step-[2], leading to $(I_*)_M$.
\item Using the gain and readout-noise (Ron) of the CCD, the noise at the $i^{th}$ pixel is calculated as $N_i= \sqrt{\frac{(I_*+Sky)_{iM}}{gain}+Ron^2_{ADU}}$.
\item Step-[5] and step-[6] lead to the estimation of the relative weight $w_i$ at each pixel of the CCD:  $w_i = (I_*)_{iM}/N_i^2$.
\item The above weights are summed up along the row: $\sum_{i=0}^n{(I_*)_{iM}}/N_i^2$.
\item Normalizing the weight $w_i$ of the pixels along the rows, we find that $W_i= \frac{(I_*)_{iM}/N_i^2}{\sum_{i=0}^n{(I_*)_{iM}}/N_i^2}$.
\item These weights are then multiplied with the signal step-[5] i.e, $W_i\times (I_*)_{iM}.$
\item Remember, the masking has led to zero values elsewhere but unity within the mask width. Therefore, a sum of the weighted signal along the row will give the total signal $S = \sum{W_i\times (I_*)_{iM}}$. This process leads to a one-dimensional extracted spectrum having a length equal to that of the CCD columns.
\item In order to calculate the total noise, we multiply the weights with the noise step-[6], i.e., $W^2_i\times N_i^2$.
\item Similar to step-[10], the total noise along the mask width is $N = \sqrt {\sum{W_i^2\times N_i^2}}$. Similar to the signal spectrum, a one-dimensional noise spectrum with a length equal to that of the CCD columns is generated.
\item Finally, we slide the mask along the rows, and for each column position, we calculate the corresponding S/N quantity. The presence of a source spectrum leads to a maximum value for S/N. The presence of two extrema (at positions $X_1$ and $X_2$) corresponds to the signature of two source spectra present on the CCD frame (see Fig.\,\ref{Fig:mask}).
\item After detecting such peaks, the above process is repeated for various mask widths in order to optimally extract the spectra ($I[X_1]$ and $I[X_2]$). These are shown in blue colour in Fig.\,\ref{Fig:mask_res}.
\item  We also extract the spectra at positions $X_1-(X_2-X_1)$ and at positions $X_2 + (X_2-X_1)$ (orange colour in Fig.\,\ref{Fig:mask_res}) and then construct the following de-contaminated spectra:
$I_D[X_1] = I[X_1] - I[X_2+(X_2 - X_1)]$ and $I_D[X_2] = I[X_2] - I[X_1-(X_2 - X_1)]$. 
\end{enumerate}
Fig.\,\ref{Fig:mask} shows two S/N peaks using the \emph{masking technique} for the corresponding spectra shown in the left panel of Fig.\,\ref{Fig:decontam}. As seen from Fig.\,\ref{Fig:mask_res}, the contaminated and decontaminated spectra look quite different.

\section{Data log and results}
\label{s:datalog}

The masking technique for removing the contamination of very closely separated quasars provides a unique method to extract the spectra of doubly imaged quasar candidates. Table\,\ref{tab:source_info_lens} lists the sources for which the spectra have been extracted in this paper and were confirmed to be lensed. Out of 57 sources, we have confirmed 11 sources to be gravitationally lensed quasars. Ten of the 11 sources are doubly imaged quasars, whereas one quasar is a quadruply imaged quasar. The spectra are provided at \url{https://github.com/PriyankaJalan14/Lens_spectra}.

\begin{table}[ht!]
\centering
\begin{minipage}{156mm}
\caption{Observation log of the gravitational lens candidates. The lensing nature is also confirmed in the quoted references.}
\label{tab:source_info_lens}
\end{minipage}
\bigskip

\begin{tabular}{ccrrrc}
\hline
\textbf{Quasar} & \textbf{Date}      & \multicolumn{3}{c}{\textbf{For each component}} & \textbf{Lens} \\
                & \textbf{Telescope} & \multicolumn{1}{c}{\textbf{RA}} & \multicolumn{1}{c}{\textbf{DEC}} & \multicolumn{1}{c}{\textbf{$m_r$}} & \textbf{confirmation} \\
\hline
0013+5119   & 10-01-2021 &  00:13:23.5   &    +51:19:05.9 &  21.43  & Probable lensed QSO \\
            & DOT        &  00:13:23.5   &    +51:19:04.6 &  20.79  & {[\citealt{Lemon2019MNRAS.483.4242L}]} \\
            &            &  00:13:23.6   &    +51:19:07.5 &  20.52  &  \\
\hline
0645$-$1929 & 10-01-2021 &  06:45:44.0   &  $-$19:29:36.6 &  21.13  & Doubly imaged QSO \\
            & DOT        &  06:45:44.1   &  $-$19:29:35.7 &  21.22  &  \\
            &            &  06:45:44.1   &  $-$19:29:37.6 &  19.10  &  \\
\hline
0803+3908   & 10-01-2021 &  08:03:57.7   &    +39:08:23.9 &  18.84  & Doubly imaged QSO \\
            & DOT        &  08:03:57.7   &    +39:08:23.1 &  19.71  &  \\
            &            &  08:03:57.8   &    +39:08:23.1 &  18.26  &  \\
\hline
0859$-$3011 & 08-04-2019 &  08:59:11.9   &  $-$30:11:34.7 &  20.23  & Doubly imaged QSO \\
            & NTT        &  08:59:11.9   &  $-$30:11:34.6 &  20.99  &  \\
            &            &  08:59:11.9   &  $-$30:11:35.4 &  20.87  &  \\
\hline
0911+0550   & 09-04-2019 &  09:11:27.6   &    +05:50:54.8 &  19.74  & Doubly imaged QSO \\
            & NTT        &  09:11:27.6   &    +05:50:53.9 &  19.50  & {[\citealt{GraL3_2019A&A...622A.165D}]} \\
\hline      
1008$-$2215 & 22-02-2020 &  10:08:53.5   &  $-$22:15:16.9 &  20.53  & Doubly imaged QSO\\
            & NTT        &  10:08:53.5   &  $-$22:15:17.9 &  20.94  &  \\
            &            &  10:08:53.6   &  $-$22:15:18.2 &  19.75  &  \\
\hline
1124+5710   & 19-03-2021 &  11:24:55.3   &    +57:10:56.6 &  18.46  & Doubly imaged QSO \\
            & DOT        &  11:24:55.5   &    +57:10:58.1 &  19.76  &  \\
\hline
1145$-$0850 & 11-02-2021 &  11:45:26.0   &  $-$08:50:06.4 &  21.68  & Probable lensed QSO \\
            & DOT        &  11:45:25.9   &  $-$08:50:07.5 &  21.28  &  \\
            &            &  11:45:24.0   &  $-$08:50:04.0 &  20.62  &  \\
\hline
1554$-$2818 & 23-02-2020 &  15:54:2.3    &  $-$28:18:36.4 &  19.49  & Doubly imaged QSO \\
            & NTT        &  15:54:2.2    &  $-$28:18:34.6 &  19.99  &  \\
            &            &  15:54:2.2    &  $-$28:18:35.6 &  19.76  &  \\
\hline
1651$-$0417 & 09-04-2019 &  16:51:04.5   &  $-$04:17:25.0 &  20.29  & Quadruply imaged QSO \\
            & NTT        &  16:51:05.5   &  $-$04:17:27.3 &  19.48  & {[\citealt{GraL6_2020arXiv201210051S}]} \\
            &            &  16:51:05.1   &  $-$04:17:27.8 &  18.98  &  \\
            &            &  16:51:05.2   &  $-$04:17:23.2 &  20.04  &  \\
\hline
1654+3318   & 13-02-2021 &  16:54:23.5   &    +33:18:02.9 &  20.82  &  Probable lensed QSO\\
            & DOT        &  16:54:23.4   &    +33:18:02.2 &  20.83  &  \\
            &            &  16:54:23.5   &    +33:18:03.1 &  20.50  &  \\
\hline       
\end{tabular}
\end{table}

 \section{Conclusions}
 \label{s:ch6_conclusions}
The Gaia GraL group aims at discovering more gravitationally lensed quasars. The challenge is to remove the spectral contamination due to the proximity of the lensed images. An optimized extraction technique is required to remove the contamination. In this paper, we have discussed the
 spectral extraction technique for the case of very nearby lensed components. We detect the high S/N peaks in
 the CCD image using a \emph{masking technique}. This technique computes the cumulative signal using a weighted sum, yielding a reliable approximation for the total counts. We then subtract the mutual spectral contamination due to the proximity of the lensed images. The width of the mask is decided through an iterative
 process. In this paper, we have efficiently extracted the spectra to
 confirm/refute 57 quasar lens candidates using this technique. Out of
 fifty-seven candidates, 11 of them are found to be lensed quasars.

\begin{acknowledgments}
This work is partially based on the observations obtained at the 3.6m Devasthal Optical Telescope (DOT) under programme ID DOT-2020-C2-P49, DOT-2020-C2-P50, DOT-2020-C2-P54, DOT-2020-C2-P59, DOT-2021-C1-P11, DOT-2021-C1-P12, DOT-2021-C1-P17, DOT-2021-C1-P33, which is a National Facility run and managed by Aryabhatta Research Institute of observational sciencES (ARIES), an autonomous Institute under the Department of Science and Technology, Government of India. The observations are also based on the European Organisation for Astronomical Research in the Southern Hemisphere under ESO programme(s) P103.A-0077, P104.A-575, P105.A-205, P106.A.215V, P108.223G, P111.24HD. PJ was supported by the Polish National Science Center through grant no. 2020/38/E/ST9/00395. This work is supported by the Belgo-Indian Network for Astronomy and astrophysics (BINA), approved by the International Division, Department of Science and Technology (DST, Govt. of India; DST/INT/BELG/P-09/2017) and the Belgian Federal Science Policy Office (BELSPO, Govt. of Belgium; BL/33/IN12).
\end{acknowledgments}

\begin{furtherinformation}

\begin{orcids}
\orcid{0000-0002-0524-5328}{Priyanka}{Jalan}
\orcid{0000-0001-5824-1040}{Vibhore}{Negi}
\orcid{0000-0002-7005-1976}{Jean}{Surdej}
\orcid{0000-0002-5074-9998}{C\'eline}{Boehm}
\orcid{0000-0003-2559-408X}{Ludovic}{Delchambre}
\orcid{0000-0002-8760-6157}{Jakob Sebastian}{den Brok}
\orcid{0000-0003-0699-7019}{Dougal}{Dobie}
\orcid{0000-0003-0228-6594}{Andrew}{Drake}%
\orcid{0000-0003-4843-8979}{Christine}{Ducourant}
\orcid{0000-0002-0603-3087}{S. George}{Djorgovski}
\orcid{0000-0002-8541-0476}{Laurent}{Galluccio}
\orcid{0000-0002-3168-0139}{Matthew J.}{Graham}
\orcid{0000-0002-3469-5133}{Jonas}{Kl\"uter}
\orcid{0000-0002-2308-6623}{Alberto}{Krone-Martins}
\orcid{0000-0003-2242-0244}{Ashish A.}{Mahabal}
\orcid{0000-0002-2686-438X}{Tara}{Murphy}
\orcid{0000-0001-6809-2536}{Anna}{Nierenberg}
\orcid{0000-0003-3739-4288}{Sergio}{Scarano}
\orcid{0000-0003-1407-6607}{Joseph}{Simon}
\orcid{0000-0003-4771-7263}{Eric}{Slezak}
\orcid{0000-0001-6116-2095}{Dominique}{Sluse}
\orcid{0000-0002-8052-7763}{Carolina}{Sp\'indola-Duarte}
\orcid{0000-0003-2686-9241}{Daniel}{Stern}
\orcid{0000-0002-6806-6626}{Ramachrisna}{Teixera}
\orcid{0000-0002-8365-7619}{Joachim}{Wambsganss}

\end{orcids}

\begin{authorcontributions}
This work results from a long-term collaboration (GAIA-GRAL) to which all authors have made significant contributions.

\end{authorcontributions}

\begin{conflictsofinterest}
The authors declare no conflict of interest.
\end{conflictsofinterest}

\end{furtherinformation}

\bibliographystyle{bullsrsl-en}

\bibliography{S10-CT02_JalanP}

\end{document}